# Complementing AC & DC Terminal Stability Analyses of MMC with Inner Loop Impedance

Chongbin Zhao, *Student Member, IEEE*, and Qirong Jiang

*Abstract*—Learning from two-level voltage source converters, the existing impedance-based stability analyses of modular multilevel converters (MMCs) primarily focus on system modes with finite closed-loop transfer functions, which consider perturbations of the current flowing into the public AC/DC terminal as the input. However, this approach may be insufficient for MMCs due to their actively controlled circulating circuit, resulting from the distributed modulation of each arm and the circulating current control (CCC). To address this limitation, two cases that are not covered by the AC/DC terminal stability analysis are initially presented to support the conjecture. Subsequently, an inner loop impedance for the circulating circuit is established, which considers the dynamics of public terminals and divides the injected voltage perturbation by the corresponding current perturbation at the same frequency. To avoid the need for a right-half plane pole check when applying the Nyquist criterion, a logarithmic derivative-based criterion is proposed to directly identify the system modes. By utilizing the inner loop impedance, it becomes possible to achieve CCC parameter tuning with stability constraints and conduct an internal stability analysis of MMC-based systems. This work provides a strong foundation for the integration of power-electronicized power systems from the perspective of classical control theories.

## I. Introduction

STABILITY analysis is a crucial aspect of designing interconnected power systems as it assesses the risk of system collapse under various disturbances [1]-[2]. With the increasing integration of power electronics devices such as two-level voltage source converters (TL-VSCs) and modular multilevel converters (MMCs), numerous harmonic stability issues have emerged in recent years [3]-[7]. The inadequate solutions for mitigating these oscillations highlight the need for a comprehensive review of commonly used stability analysis theories.

### A. Motivations

Transfer function (TF)-based and state space (SS)-based methods are two commonly employed approaches for harmonic stability analysis. In practical scenarios, converter manufacturers often provide transmission system operators (TSOs) with black/gray-box models that lack detailed control parameters. As a result, constructing a high-fidelity SS for an actual system becomes impractical [8]. Instead, impedance/admittance models (IMs/AMs) are defined using a set of voltage and current perturbations. By utilizing frequency responses rather than intricate IM expressions, stability analysis becomes feasible [9]-[10]. Since voltage and current are fundamental quantities in circuit analysis and are present in every component of the system, impedance-based methods have found extensive real-world applications.

Among the modeling techniques for an AC IM, sequence IMs are well-suited for system-level analysis and can be easily validated through field tests. This is because the terminal phase can be inherently modeled within sequence IMs. On the other hand, if $dq$ IMs are adopted, the utilization of multi-$dq$ transformations and synchronizations becomes unavoidable [10]. Furthermore, the *terminal* IM/AM Furthermore, the terminal IM/AM is a multiple-input multiple-output (MIMO) system [11] considering the frequency coupling. It is also possible to transform the terminal IMs into single-input single-output (SISO) loop IMs which can reflect the partial system modes. This transformation involves solving the Schur complement of a submatrix of the terminal impedance sum [12]. While frequency responses of IMs can be obtained directly through multiharmonic linearization in the frequency domain [13], [14], they can also be derived based on the relationship between TF and SS. This raises a fundamental question from the perspective of classical control theories [15]: For a complex system that consists of multiple physical circuits, such as MMCs, how many loop IMs are required to ensure the *internal stability analysis* [15]? In other words, can the selected set of loop IMs cover all the system modes considering their inconsistent zero-pole cancellations?

The objective of this work is to uncover and address the inherent limitations of using commonly employed terminal AC & DC loop IMs to assess the internal stability analysis of MMC-based systems.

### B. Related Works

The lack of attention on internal stability analysis using IMs for TL-VSCs has been widely reported. In [16], it is confirmed that the commonly used active damping for an LCL-type VSC can only cancel the resonant poles of the single TF involving the sampling current. However, due to the introduction of three state variables by the LCL filter, a pair of critical stable modes can be factored out from the TFs involving the non-sampling state variables. These modes allow the characteristic harmonic to flow through a physical circuit and trigger the protection logic. Another observation in [10] involves antiphase current

This work was supported in part by the National Natural Science Foundation of China under Grant U22B6008. *(Corresponding author: Qirong Jiang)*
The authors are with the Department of Electrical Engineering, Tsinghua University, Beijing 100084, China (e-mail: zhaocb19@mails.tsinghua.edu.cn; qrjiang@mail.tsinghua.edu.cn).

2harmonics between two VSCs in a tree-like power supply. It is believed that the negative-damping mode can only be identified by selecting a partition point in the branch, excluding the public terminal, and using the corresponding IMs. This preference can also be extended to aggregated wind farms. For instance, if the DC-bus voltage controllers are not well-tuned and induce DC resonances, the resulting harmonics may propagate to the AC side. However, they may not be revealed by the impedance-based AC stability analysis [17], [18] because the negative-damping modes are canceled in the corresponding loop IMs. In a word, to ensure a more reliable impedance-based internal stability analysis, it is crucial to form sufficient IMs that cover each physical circuit, thereby aiding in locating oscillations.

Drawing from the experience of TL-VSCs, the initial establishment of 2×2 AC terminal IMs followed by 1×1 DC terminal IMs for MMCs has been adopted [4]-[6], [14], [18]-[22]. The Nyquist criteria are then utilized to assess stability using the loop gain formed by either terminal or loop IMs. In comparison to other technical approaches, the superiority of sequence IMs based on multiharmonic linearization is further highlighted. This is because a) sequence loop IMs possess intuitive physical meaning, as perturbations can be directly added to the total 6 physical circuits combined by the positive-/negative-/zero-sequence (PS/NS/ZS) and the common-/differential-mode (CM/DM) circuits, and b) the computational burden is relatively low, as only the perturbation of one arm/phase is used, and the irrational form of pure time delay can be maintained in the IM [18]. Furthermore, the internal dynamics of MMCs, resulting from the instantaneous unbalance of arm voltages and circulating current control (CCC), reveal significant effects in the IMs concerning interconnected circuits. However, these effects in the existing AC & DC IM modeling of MMCs should be regarded as indirect effects that act on the off-diagonal block submatrices when obtaining the Schur complement. It can be inferred that the dedicated IM for the inner circulating circuit (i.e., the AC-CM circuit) is required for the internal stability analysis of MMC. The conjecture mentioned here can be related to the former discussion in [17], [18]: Even if DC-CM dynamics are considered in the AC-DM impedance modeling to improve the accuracy of AC stability analysis, performing DC stability analysis is still necessary to complement the canceled mode in the AC stability analysis.

In the conventional implementation of the impedance-based method, the open-loop gains are free from right-half plane (RHP) poles when the source/load subsystem can stably operate [15]. However, manually splitting the loop IM into two subsystems to form the equivalent open-loop gain lacks empirical evidence supporting the absence of RHP poles because these subsystems do not physically exist and cannot be validated through simulations. When a clockwise encirclement around the origin occurs, the stability assessment becomes ambiguous, necessitating a reliable RHP pole check using frequency responses. However, the existing methods [23], [24] are complex and not straightforward. The practical criterion proposed in [25] aims at directly identifying the system mode, which uncovers the essence of the stability analysis by directly identifying the RHP zero of the closed-loop gain. However, it has been noted in [18] that the criterion may erroneously classify a stable mode as unstable due to the obfuscation of polynomials except for the focused pair of zeros. Hence, a novel criterion that can equally treat each zero and pole is needed to better cooperate with the loop IMs.

### C. Technical Contributions and Paper Structure

This work contributes and organizes the following aspects:
a) In Section II, the studied MMC-based system is introduced, and typical cases are presented where AC/DC terminal stability analysis falls short. This highlights the limitations of existing studies.
b) In Section III, leveraging the AC/DC terminal dynamics in a black-box manner that are offered by TSOs, an intuitive derivation of inner loop IM based on multiharmonic linearization is achieved, which can be easily implemented by MMC manufacturers.
c) In Section IV, the logarithmic derivative-based stability criterion using the frequency responses of loop IMs is proposed. It serves as a basis to identify unstable modes through either numerical estimation or accurate unconstrained optimization.

Section V validates the proposed methods and Section VI gives the discussions and the conclusions.

## II. THE STUDIED CASES

### A. Basics of the MMC-BTB System

The MMC-based back-to-back (MMC-BTB) system under study is depicted in Fig. 1. In comparison to the previous work [18], the ideal transformer models are configured at both sending-end (SE) and receiving-end (RE) to isolate the ZS-DM current. Additionally, the voltage level and key parameters have been updated in Table I to align with an actual asynchronous grid interconnection system [6].

Most of the control loops, including grid-following synchronization, $dq$ decoupled per-unit value control, outer loop control modes (both ends regulate reactive power), and the structure of inner current control (comprising fundamental frequency PS/NS current control & double fundamental frequency CCC), can be learned from [18]. However, it is worth emphasizing the two reference generations of CCC that correspond to the two cases studied in this work, as shown in Fig. 2:

a) The first CCC enforces the $d$ & $q$ references $i^{C*}_{d/qpu}$ to be 0, which is referred to as circulating current suppressing control (CCSC).

b) For the second CCC, an outer loop is introduced, defined as forced CCC (FCCC), which aims to suppress the 2$^{nd}$-order components of each arm voltage to 0.

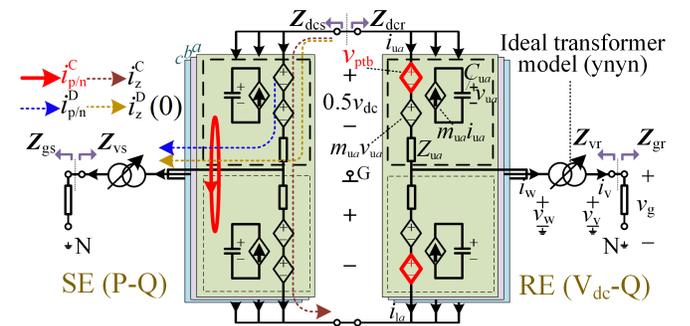

Fig. 1. Diagram of the MMC-BTB system.

TABLE I
KEY PARAMETERS OF THE SYSTEM

| Categories | Symbols | Value (Unit) |
|---|---|---|
| Base | $S_b, V_{acb}, V_{dcb}, f_1$ | 1380 MVA, 525 kV, 840 kV, 50 Hz |
| Grid | $v_{gr}, v_{gs}$ | 542.3∠82°, 620.2∠90° |
|  | $L_{gr}, L_{gs}, R_{gr}, R_{gs}$ | 100 mH, 0, 0, 40 Ω |
| MMC | $L, R$ | 140 mH, 1Ω, 1000 |
|  | $C', N$ | 11000 μF, 500 |
| Transformer | $a, X_T$ | 1.2007, 0.14 pu |
| Controllers ($h$==$K_p$ +1/$sT_i$) | $P_s^*, Q_s^*/Q_r^*, v_{dcr}^*$ | 1250 MW, 0/200 MVA, 840 kV |
|  | $h_{PQ}(s), h_{ov}(s), h_{v_{dc}}(s)$ | 1+1/0.01$s$, 1.3+1/0.01$s$, 10+1/0.05$s$ |
|  | $h_{PLL}(s)$ | 100+1/0.05$s$ |
|  | $h_{i1}(s), h_{i2}(s)$ | 0.35+1/0.1$s$, 0.8+1/0.01$s$ |
|  | $T_{dr}, T_{ds}$ | 460μs, 0μs |

It should be noted that grid synchronization is considered while decoupling and feedforward compensation are not taken into account for the CCCs.

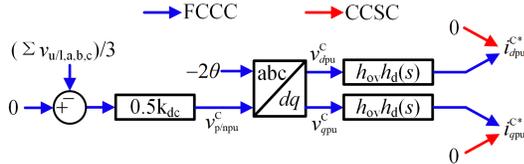

Fig.2. Outer loop control (inner loop reference generation) of two CCCs.

## B. Simulation Results in PSCAD

Three tests are conducted for the CCSC case, as shown in Fig. 3 (located at the bottom of this page). The tests involve stepping the $K_{pi2}$s for both ends from 0.8 to 0.1 (Fig. 3(a)), 1.9 (Fig. 3(b)), and 2.2 (Fig. 3(c)) at $t$=5 s. For simplicity, only the DC variable of the active power of the RE MMC ($P_r$) is displayed, which indicates a transition from stability to instability in all three tests. By extracting the data from a quasi-linear oscillation interval $T_i$ (0.29 s in Fig. 3(a), 0.79 s in Fig. 3(b), and 0.07 s in Fig. 3(c)), the dominant initial oscillation frequencies of $P_r$ are determined as 70, 150, and 600 Hz, respectively. It is worth noting that in Fig. 3(a) and(c), the primary 66/606 Hz harmonic diverges much faster than the secondary 176/150 Hz harmonic. The sustained oscillation frequencies can provide more practical approximations to initial oscillation frequencies, even if they should not be technically equal. The divergence rate ($\alpha$) of $P_r$ is calculated using the following formula:

$$\alpha=\ln[P_r(t_0+T_i)/P_r(t_0)]/T_i \quad (1)$$

where $t_0$ is the starting time of sampling. Therefore, for the CCSC case, the divergence rates of the three tests are calculated as 2.20, 0.51, and 2.99 s$^{-1}$, respectively.

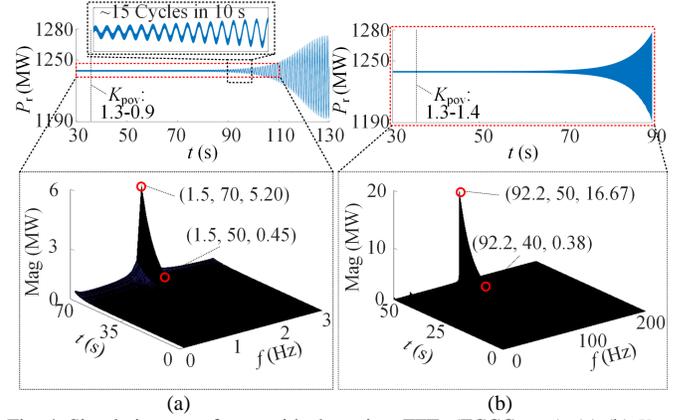

Fig. 4. Simulation waveforms with short-time FFTs (FCCC case). (a)-(b) $K_{p\_ov}$ changes from 1.3 to 0.9/1.4 at $t$=35 s, respectively.

The same pattern is also applied to the FCCC case as shown in Fig. 4, where $K_{p\_ov}$s steps from 1.3 to 0.9 (Fig. 4(a)) or 1.4 (Fig. 4(b)) for both ends at $t$=35 s, and the dominant initial oscillation frequencies are 1.5 and 92.2 Hz while the divergence rates are 0.12 and 0.37 s$^{-1}$ for the two tests, respectively.

## C. Nyquist Plots for AC/DC Terminal Stability Analyses

The stability analyses using Nyquist plots are shown in Fig. 5. Each subfigure represents a different loop IM involving the AC-DM-RE, AC-DM-SE, and DC-CM circuits, respectively, from left to right. The eigenloci are obtained by substituting frequencies ranging from –2900 Hz to 3000 Hz into the equivalent open-loop gains, as defined in [18]. It is evident that tuning $K_{\text{pi2}}/K_{\text{pov}}$ has minimal effect on the nature of each Nyquist plot, and no encirclement around the origin (–1, 0j) is observed. According to the Nyquist criterion [15], the absence of encirclement indicates that the negative-damping modes of the unstable interconnected systems are canceled in the three loop IMs, despite the presence of RHP poles in some open-loop gains due to the instability of certain subsystems when fed by ideal voltage/current sources. This observation highlights the need for a new closed-loop TF, namely the inner loop IM.

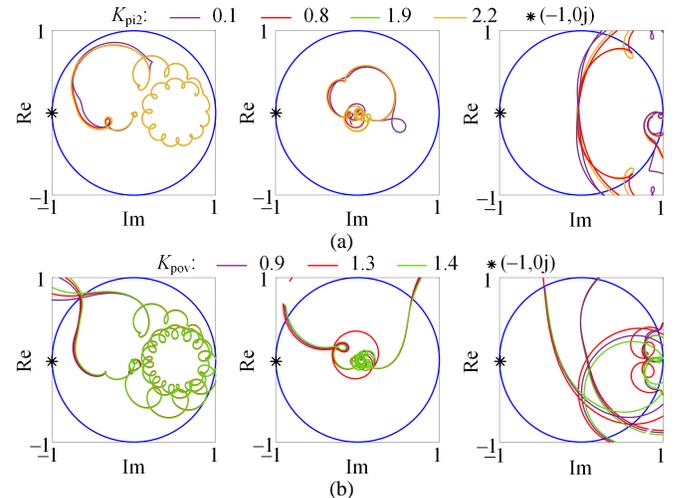

Fig. 5. Nyquist plots for terminal stability analyses. (a)-(b) CCSC/FCCC case.

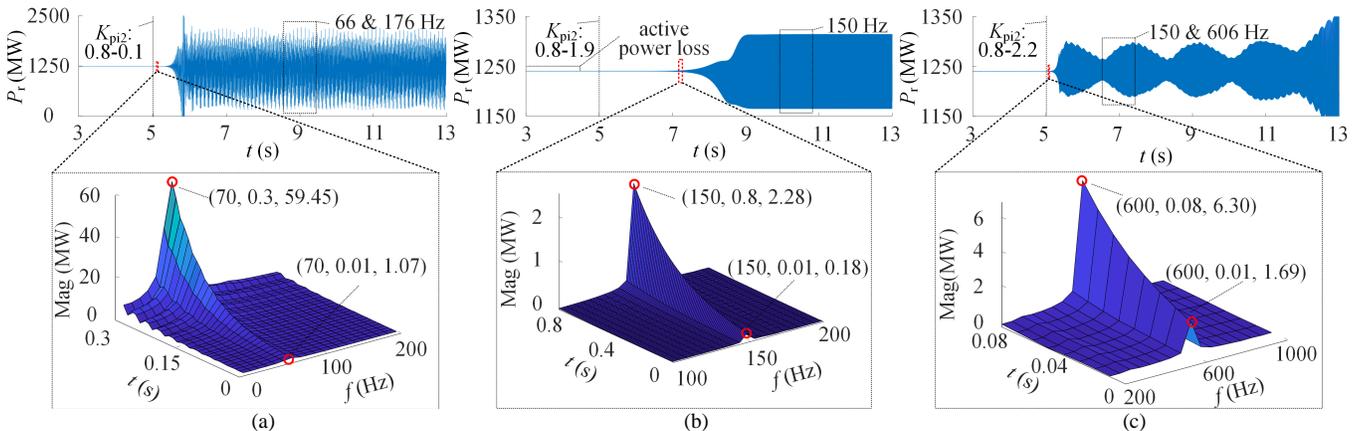

Fig. 3. Simulation waveforms with short-time FFTs (CCSC case). (a)-(c) $K_{pi2}$ changes from 0.8 to 0.1/1.9/2.2 at $t$=35 s, respectively.

## III. Inner Loop IM Derivation of MMCs

### A. Steady-State Harmonic Calculation

The calculation of steady-state harmonics in an assumed stable system is a crucial aspect that has not received sufficient attention in the impedance-based method until the recent work presented in [18]. In this subsection, the principles are introduced related to the newly adopted FCCC in this work.

Eq. (2) represents a set of time-domain expressions for the power stage, which hold for the upper arm of phase A of an MMC. The symbols used in (2) can be understood by referring to Fig. 1 & Table I, and are omitted for the sake of brevity.

$$\begin{cases} L(di(t)/dt) + Ri(t) = 0.5v_{\text{dc}}(t) - v_{\text{w}}(t) - m(t)v(t) - v_{\text{ptb}}(t) \\ C(dv(t)/dt) = m(t)i(t), \ C = C'/N \\ v_{\text{w}}(t) = \{v_{\text{v}}(t) + L_{\text{x}}[d(2i^D_{\text{p/n}}(t)/k_{\text{a}})/dt]\}/k_{\text{a}} \end{cases} \quad (2)$$

By transforming (2) from the time domain to the frequency domain, a new set of equations is obtained:

$$\begin{cases} \boldsymbol{i} = \boldsymbol{Y}_{\text{LR}}(0.5\boldsymbol{v}_{\text{dc}} - \boldsymbol{v}_{\text{v}}/k_{\text{a}} - \boldsymbol{M}\boldsymbol{v} - \boldsymbol{v}_{\text{ptb}}) \\ \boldsymbol{Y}_C\boldsymbol{v} = \boldsymbol{M}\boldsymbol{i} \end{cases} \quad (3)$$

where the lowercase bold symbol represents a frequency domain vector formed by the Fourier series coefficients of the corresponding time domain signal in (2), while the capitalized bold symbol represents either the Toeplitz matrix based on the conjugate symmetrical vector or the frequency-dependent diagonal matrix based on $L$, $C$, and $R$. Considering the non-infinite harmonic coupling due to the MMC modulation, it is practical to truncate the harmonics at a certain order n. This means that the order of each matrix is $(2n+1)\times(2n+1)$, while the order of each vector is $(2n+1)\times1$. In this work, n is selected as 4. Taking the arm current vector $\boldsymbol{i}$ as an example, it can be represented as:

$$\boldsymbol{i} = [i\langle-4\rangle, i\langle-3\rangle, \cdots, i\langle0\rangle, \cdots, i\langle3\rangle, i\langle4\rangle]^T \quad (4)$$

The approach for solving the steady-state harmonics involves formulating a set of nonlinear equations derived from (3) and using the Newton-Raphson iteration to solve for the unknown complex coefficients in each vector. The control stage, which determines the insertion index vector $\boldsymbol{m}$, also introduces additional equations and influences the calculation results, particularly through the outer loop control references. For instance, the active/reactive power references act as constraints for the power flow calculation at each end, which involves $\boldsymbol{i}$ and $\boldsymbol{v}_{\text{v}}$ to construct the equations. Compared with the CCSC case, the targeted variables for the FCCC case change because $v\langle2\rangle$ instead of $i\langle2\rangle$ is known as 0+0j, but the number of equations remains the same.

The steady-state calculations are compared in Table II, highlighting the overall differences between the two cases. Notably, it demonstrates the decrease in the root mean square value of $v$ instead of $i$ for the FCCC case compared with the CCSC case, which satisfies the intuitive cognitions.

TABLE II
Comparison of Steady-State Harmonics

| Variable | RE | | SE | |
|---|---|---|---|---|
| | CCSC | FCCC | CCSC | FCCC |
| $m\langle1\rangle$ | –0.2070 –0.0382j | –0.2050 –0.0442j | –0.2115 0.0393j | –0.2104 –0.0464j |
| $m\langle2\rangle$ | 0.0040 –0.0116j | 0.0126 0.0276j | 0.0015 0.0124j | 0.0081 0.0299j |
| $i\langle0\rangle$ | 0.4938 | 0.4935 | –0.4938 | –0.4935 |
| $i\langle1\rangle$ | 0.5789 0.0955j | 0.5786 0.0965j | –0.5836 0j | –0.5836 0j |
| $i\langle2\rangle$ | 0 0j | 0.2163 0.1178j | 0 0j | –0.2378 0.0834j |
| $v\langle0\rangle$ | 837.9021 | 834.7172 | 833.8890 | 831.5551 |
| $v\langle1\rangle$ | 3.1588 –27.2706j | –0.7698 –20.7151j | –3.8594 27.2314j | –6.7791 –20.0770j |
| $v\langle2\rangle$ | –3.4404 8.2626j | 0 0j | –2.1049 –8.8796j | 0 0j |
| $v\langle3\rangle$ | –0.3138 –0.1703j | –2.3610 1.4321j | –0.3611 0.0432j | –2.2413 2.0301j |

### B. Inner Loop IM Derivation

Multiharmonic linearization of (3) for each MMC results in:

$$\begin{cases} \Delta\boldsymbol{i} = \boldsymbol{Y}_{\text{LR}}(0.5\Delta\boldsymbol{v}_{\text{dc}} - \Delta\boldsymbol{v}_{\text{v}}/k_{\text{a}} - \boldsymbol{M}\Delta\boldsymbol{v} - \boldsymbol{V}\Delta\boldsymbol{m} - \Delta\boldsymbol{v}_{\text{ptb}}) \\ \boldsymbol{Y}_C\Delta\boldsymbol{v} = \boldsymbol{M}\Delta\boldsymbol{i} + \boldsymbol{I}\Delta\boldsymbol{m} \end{cases} \quad (5)$$

where the prefix $\Delta$ represents the perturbation. Since the inner loop IM connects $\Delta\boldsymbol{v}_{\text{ptb}}$ with $\Delta\boldsymbol{i}$, the vectors $\Delta\boldsymbol{v}_{\text{v}}$, $\Delta\boldsymbol{v}_{\text{dc}}$, $\Delta\boldsymbol{v}$, and $\Delta\boldsymbol{m}$ must be eliminated, so the extra three equations are needed:

$$\Delta\boldsymbol{v}_{\text{dc}} = -3\hat{\boldsymbol{Z}}_{\text{dc}}\Delta\boldsymbol{i}, \ \Delta\boldsymbol{v}_{\text{v}} = (2/k_{\text{a}})\boldsymbol{Z}_{\text{g}}\Delta\boldsymbol{i},$$
$$\Delta\boldsymbol{m} = \boldsymbol{B}_i\Delta\boldsymbol{i} + \boldsymbol{B}_{v_{\text{v}}}\Delta\boldsymbol{v}_{\text{v}} + \boldsymbol{B}_{v_{\text{dc}}}\Delta\boldsymbol{v}_{\text{dc}} + \boldsymbol{B}_v\Delta\boldsymbol{v} \quad (6)$$

By substituting (6) into (5) and rearranging the equations, the concerned inner loop IM $Z^C_{\text{n/p\_L}}(s)$ can be inverted from the specific element of the TF matrix $\boldsymbol{Y}^{\text{C\_ptb}}_{\text{AC}}$:

$$\Delta\boldsymbol{i} = \boldsymbol{Y}^{\text{C\_ptb}}_{\text{AC}}\Delta\boldsymbol{v}_{\text{ptb}} \Rightarrow Y^C_{\text{n/p\_L}}(s) = \boldsymbol{Y}^{\text{C\_ptb}}_{\text{AC}}{}_{(n\pm2,n\pm2)}, \ Z^C_{\text{n/p\_L}}(s) = Y^{C}_{\text{n/p\_L}}{}^{-1}(s),$$
$$\boldsymbol{Y}^{\text{C\_ptb}}_{\text{AC}} = -[(\boldsymbol{A}+\boldsymbol{B}\boldsymbol{B}_i) + \underbrace{(\boldsymbol{Y}_C\boldsymbol{Y}_{\text{LR}}+\boldsymbol{B}\boldsymbol{B}_{v_{\text{ac}}})(2/k_{\text{a}}^2)\boldsymbol{Z}_{\text{g}}}_{\text{AC terminal dynamics}} + \underbrace{(\boldsymbol{Y}_C\boldsymbol{Y}_{\text{LR}}-\boldsymbol{B}\boldsymbol{B}_{v_{\text{dc}}})\hat{\boldsymbol{Z}}_{\text{dc}}}_{\text{DC terminal dynamics}}]^{-1}. \quad (7)$$

where

$$\begin{cases} \boldsymbol{A} = (\boldsymbol{Y}_C - \boldsymbol{I}\boldsymbol{B}_v) + \boldsymbol{Y}_{\text{LR}}\boldsymbol{M}(\boldsymbol{M}+\boldsymbol{V}\boldsymbol{B}_v) \\ \boldsymbol{B} = \boldsymbol{Y}_{\text{LR}}[(\boldsymbol{M}+\boldsymbol{V}\boldsymbol{B}_v)\boldsymbol{I} + (\boldsymbol{Y}_C - \boldsymbol{I}\boldsymbol{B}_v)\boldsymbol{V}] \end{cases} \quad (8)$$

For the CCSC case, $\boldsymbol{B}_v$ is a zero matrix, and the remaining TF matrices in (6) can be derived following the approach described in [18]. However, for the FCCC case, $\boldsymbol{B}_{v_{\text{v}}}$ needs to be adjusted due to the nonzero circulating current which involves the (inverse) Park transformation. This adjustment accounts for the additional dynamics of the phase-locked loop and the perturbation in the terminal AC voltage. Correspondingly, although $v(t)$ is the input of Park transformation in Fig. 2, it will not contribute an extra term in $\boldsymbol{B}_{v_{\text{v}}}$ because $v\langle2\rangle$ is regulated to 0. On the other hand, $\boldsymbol{B}_v$ is a diagonal matrix and follows a similar principle as CCSC contributes to $\boldsymbol{B}_i$:

$$\boldsymbol{B}_v(s)(\text{x,x}) = 0.5(1-g)k_m k_{\text{dc}} \cdot h_{\text{d}}(s) \cdot \{[h_{\text{i2}} \cdot h_{\text{ov}}](s+j(x-2g)\omega_1)\},$$
$$g = 2\text{mod}[(x-1-n),2]-1, \ x = n\pm2. \quad (9)$$

### C. Remarks

a) Eq. (5) indicates that the AC-DM or DC-CM loop IM can be easily determined by extracting the element at the position of $(n\pm1, n\pm1)$ or $(n, n)$ from $\boldsymbol{Y}^{\text{C\_ptb}}_{\text{AC}}$ and multiplying $k_{\text{ac}}$ or $k_{\text{dc}}$ (1/3 or $a/2$, and the scaling factors do not affect the identified modes).

It is important to note that the superficial differences in the treatment of terminal AC/DC dynamics make it inappropriate to consider the neglected terms in [18] as the accurate inner loop IMs in this work.

b) The black-box treatment of $\mathbf{Z}_g$ and $\hat{\mathbf{Z}}_{dc}$ in (6) is feasible because only the frequency responses of the loop IMs will be utilized for stability assessments: TSOs can collect the frequency responses of each AC/DC terminal IM and provide this information to MMC manufacturers. The manufacturers can then assess whether a negative-damping issue can be observed using $Z^C_{n/p\_L}(s)$, which is available for engineering purposes [5], [6].

c) Unlike the AC/DC terminal IMs, it is impractical to perform field tests on $Z^C_{n/p\_L}(s)$ once the MMC is constructed. Therefore, theoretical analysis becomes necessary to understand the behavior of $Z^C_{n/p\_L}(s)$ and its implications for stability assessment.

## IV. LOGARITHMIC DERIVATIVE-BASED STABILITY CRITERION

### A. Logarithmic Derivative with Re-Im Parts Separation

Similar to the conjugate symmetry observed between PS & NS-DM loop IMs [12], $Z^C_{p\_L}(s)$ and $Z^C_{n\_L}(s)$ also exhibit conjugate symmetry, indicating that they reflect the same mode behavior. Taking $Z^C_{n\_L}(s)$ as an example, it can be expressed in a factored pole-zero form:

$$Z^C_{n\_L}(s) = N(s)/D(s) = \prod_{i=1}^{i_z} a_{zi}(s-z_i) \Big/ \prod_{i=1}^{i_p} a_{pi}(s-p_i) \quad (10)$$

where $a_{z/p}$, $z$, and $p$ represent the flat gain, the zero, and the pole, while $i_{z/p}$ is the number of roots of the numerator/denominator polynomial $N(s)/D(s)$, respectively.

Since performing an overall calculation on (10) is intractable, the basic unit of a first-order polynomial (by substituting $s=j\omega$) $g_z(\omega)=a_z(j\omega_z-\lambda_z)$, is focused on, where the complex $\lambda_z$ is the system mode of interest, taking the form of $\lambda_z=\alpha_z+j\omega_z$. Here, $\alpha_z$ and $\omega_z$ ($=2\pi f_z$) are both real numbers that respectively represent the theoretical divergence rate and initial oscillation frequency associated with the mode. The logarithmic derivative ($D_L(\cdot)$) of $g_z(\omega)$ is defined as:

$$\begin{aligned} D_L(g_z(\omega)) &= d\log(g_z(\omega))/d\omega = d(g_z(\omega))/(g_z(\omega)d\omega) \\ &= j/(j\omega-\lambda_z) = j/[-\alpha_z+j(\omega-\omega_z)] \end{aligned} \quad (11)$$

It is found that $a_z$ is excluded in (11). Separating the real and imaginary parts (Re-Im) of (11) yields two real functions:

$$\begin{cases} \text{Re}[D_L(g_z(\omega))] = (\omega-\omega_z)/[(\omega-\omega_z)^2+\alpha_z^2] \\ \text{Im}[D_L(g_z(\omega))] = -\alpha_z/[(\omega-\omega_z)^2+\alpha_z^2] \end{cases} \quad (12)$$
$$\Rightarrow \text{Re}[D_L(g_z(\omega))]|_{\omega=\omega_z}=0, \ \text{Im}[D_L(g_z(\omega))]|_{\omega=\omega_z}=-1/\alpha_z.$$

Continuing to differentiate $\text{Re}(\cdot)$ and $\text{Im}(\cdot)$ in (12) yields:

$$\frac{d\{\text{Re}[D_L(g_z(\omega))]\}}{d\omega}\Big|_{\omega=\omega_z} = \frac{1}{\alpha_z^2}, \ \frac{d\{\text{Im}[D_L(g_z(\omega))]\}}{d\omega}\Big|_{\omega=\omega_z} = 0,$$
$$\frac{d^2\{\text{Re}[D_L(g_z(\omega))]\}}{d\omega^2}\Big|_{\omega=\omega_z} = 0, \ \frac{d^2\{\text{Im}[D_L(g_z(\omega))]\}}{d\omega^2}\Big|_{\omega=\omega_z} = \frac{2}{\alpha_z^3}. \quad (13)$$

Eqs. (12) and (13) reveal that at $\omega=\omega_z$, $\text{Im}[D_L(g_z)]$ exhibits a negative minimum ($\alpha_z>0$) or a positive maximum ($\alpha_z<0$), while $\text{Re}[D_L(g_z)]$ exhibits a definite positive slope. The presence of quadratic terms in $\text{Re}(\cdot)$ and $\text{Im}(\cdot)$ of (12) leads to a rapid decay of function values to zero as $\omega$ deviates from $\omega_z$. This behavior can be regarded as a characteristic of *frequency selectivity*.

Thanks to the logarithmic operation of $D_L(\cdot)$, the conclusions regarding $g_z(s)$ can be readily extended to $Z^C_{n\_L}(s)$:

$$D_L(Z^C_{n\_L}(s)) = \sum_{i=1}^{i_z} D_L(N(s)) + \sum_{i=1}^{i_p} -D_L(D(s)) \quad (14)$$

In (14), the reason for using a minus sign to multiply $D_L(D(s))$ is that $D(s)$ is the divisor of $Z^C_{n\_L}(s)$. This leads to the opposite property of Re-Im in (12) with the 1st- and 2nd-order derivative in (13) at $\omega=\omega_z$ for the basic unit of $D(\omega)$ compared to $g_z(\omega)$. Hence, zeros and poles of a TF can be distinguished by $D_L(\cdot)$.

### B. Stability Criterion

Suppose $f_p$ is the PS-DM (not NS-CM) perturbation frequency [14] and $f_1$ is the fundamental frequency. Set $f_{p0}$ ($f_{p0}>f_1$) and let $f_p$s ranges from $2f_1-\max(f_{p0})$ to $\max(f_{p0})$ with a small-enough step (such as 0.01 Hz). Since partial system modes can be factored out from $N(s)$ of $Z^C_{n\_L}(s)$, to determine whether a negative-damping mode with $\alpha_z>0$ potentially exists, the following process can be employed:

a) Obtain the frequency response of $Z^C_{n\_L}(s)$ by substituting $s=2\pi f_p$ in each TF matrix of (7) and (8) and performing the matrix operations.

b) Calculate $D_L(Z^C_{n\_L}(s))$ based on the *difference method* as the last equation of the first line of (11) reveals, then separating the Re-Im parts of the calculated $D_L(Z^C_{n\_L}(s))$;

c) Locate an unstable mode within the frequency range where a minimum of $\text{Im}[D_L(Z^C_{n\_L}(s))]$ coexists with a positive slope of $\text{Re}[D_L(Z^C_{n\_L}(s))]$.

Based on the derivation of $D_L(\cdot)$, $\text{Im}[D_L(\cdot)]$, and $\text{Re}[D_L(\cdot)]$ in (11) and (12), the proposed stability criterion is insensitive to $\omega_z$. This means that substituting $s=2\pi n f_1 j$ into $Z^C_{n\_L}(s)$ will not lead to a special case. When a critical stable mode with $\alpha_z=0$ can be factored out from $D(s)$ of $Z^C_{n\_L}(s)$, applying L'Hospital's rule in (12) causes $\text{Im}[D_L(Z_{cm}(s))]$ to be infinite. Hence, the proposed criterion can identify this feature and remains valid.

### C. Mode Identification

As indicated in the previous subsection, the proposed criterion is primarily used for positioning an unstable mode. In actual mode identification, which is of concern to TSOs, the two cases involving the negative damping mode as shown in Fig. 6 are commonly observed, which yields two methods.

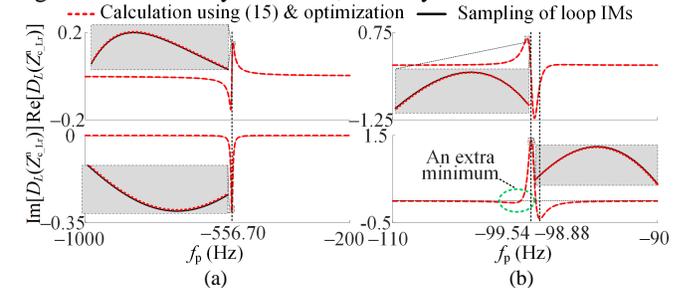

Fig. 6. Mode identification. (a) Single zero case. (b) Zero-pole pair case.

Fig. 6 (a) corresponds to the single zero case ensured by the ideal frequency selectivity, i.e., only a single zero exists over a wide frequency range, resulting in a minimum of $\text{Im}[D_L(Z^C_{n\_L}(s))]$ with the positive slope of $\text{Re}[D_L(Z^C_{n\_L}(s))]$ at the same frequency. Therefore, the corresponding frequency can be identified as $f_z$, and $\alpha_z$ can be calculated using (12). Such a process is essentially a numerical estimation.

Fig. 6 (b) corresponds to the zero-pole pair case, where a zero and a pair coexist in a very close frequency range. This

<pre>leads to two minima and one maximum of $\text{Im}[D_L(Z^C_{n\_L}(s))]$ (note that the two minima do not correspond to two modes). In such a case, the ideal frequency selectivity is disrupted, and interactions occur between the zero-pole pair makes numerical estimation inaccurate. To address this issue, a new approach based on unconstrained optimization is proposed. According to (12) and (14), the specific $\text{TF}_{\text{cal}}$ that can output the frequency responses of $\text{Im}(\cdot)$ and $\text{Re}(\cdot)$ can be written as:</pre>

$$\begin{cases} \text{TF}^{\text{Re}}_{\text{cal}}(x,\omega) = \dfrac{(\omega-\omega_z)}{(\omega-\omega_z)^2+\alpha_z^2} - \dfrac{(\omega-\omega_p)}{(\omega-\omega_z)^2+\alpha_z^2} \\ \text{TF}^{\text{Im}}_{\text{cal}}(x,\omega) = \dfrac{-\alpha_z}{(\omega-\omega_z)^2+\alpha_z^2} - \dfrac{-\alpha_p}{(\omega-\omega_p)^2+\alpha_p^2} \end{cases}, x=\{\omega_z,\alpha_z,\omega_p,\alpha_p\}. \quad (15)$$

The next step involves sampling sufficient points from Fig. 6 (b). The sampling rate is set to 0.1 Hz, resulting in $n_t$ groups of Re-Im data. By substituting $\omega=\omega_{\text{sam}}$ (the frequency labels) into (15), a series of frequency responses are calculated. The objective of mode identification is to ensure that the calculated values closely match the sample values for both $\text{Im}(\cdot)$ and $\text{Re}(\cdot)$. To achieve this, a nonlinear unconstrained multivariable function $f(x)$ can be constructed:

$$f(x) = \sum_{n=1}^{n_t}\{[\text{TF}^{\text{Re}}_{\text{cal}}(x,n)-\text{TF}^{\text{Re}}_{\text{sam}}(n)]^2 + [\text{TF}^{\text{Im}}_{\text{cal}}(x,n)-\text{TF}^{\text{Im}}_{\text{sam}}(n)]^2\} \quad (16)$$

The optimization-based mode identification involves finding a local minimum of the function f(x) using derivative-free optimization (such as the simplex method using MATLAB command *fminsearch*) or gradient optimization (such as the quasi-Newton or trust-region method using MATLAB's command *fminunc*). The numerical estimation results can be used as the initial values for the optimization process. This approach can be applied to both the single zero case and other more complex cases by adjusting equation (15) accordingly. The accuracy of the optimization-based mode identification is reflected in Fig. 6, showing the high precision achieved in both cases.

### D. Remarks

It is important to note that the unconstrained optimization approach is more versatile compared to common regression methods like vector fitting [26] for identifying specific unstable modes in stability analyses. There are two main reasons for this:

a) Vector fitting assumes the targeted TF to be a rational polynomial, but the sequence loop IM is an irrational polynomial [11]. Therefore, vector fitting is not suitable for identifying sequence loop IMs and is impractical in this context.

b) Even if a rational polynomial (e.g., the $dq$ IM) with a nature of self-symmetry is established, the proposed focuses solely on the negative-damping modes of interest, rather than considering the flat gains, zeros, and poles.

Both methods require a predefinition of zero-pole orders, which can involve some trial and error in the implementation. However, the unconstrained optimization method still offers advantages in terms of universality, efficiency, and intuitive identification of negative-damping modes.

## V. VALIDATION OF THE PROPOSED METHODS

### A. Validation of Inner Loop IM

In the frequency scans conducted using PSCAD, a set of co-phase NS-CM voltage perturbations at $f_p+f_1$ is added in the upper and lower arms of each MMC as highlighted in red in Fig. 1. The perturbations in the circulating circuit at the same frequency are then sampled to obtain the frequency scan results. In Fig. 7, for the CCSC case, the frequency responses of $Z^c_{n\_L}(s)$ with a 1 Hz frequency step (solid lines) are compared with the frequency scans (marks). The results demonstrate a high level of consistency across the wideband frequency range, confirming the accuracy of the theoretical derivations of the inner loop IMs.

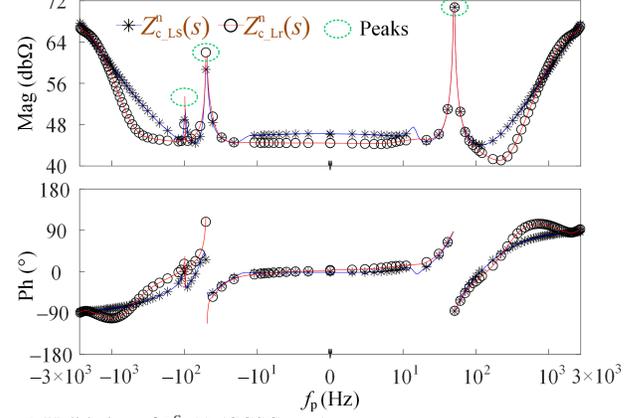

Fig. 7. Validation of $Z^c_{n\_L}(s)$ (CCSC case).

According to Table I, a pure time delay of 460 $\mu$s is added to the control loop of the RE MMC, while it is not present in the SE MMC. This difference in time delay results in discrepancies in the two $Z^c_{n\_L}(s)$s above $2f_1$ and below $-2f_1$, where the inner loop current control with time delay has a significant impact on the frequency responses. Correspondingly, since the elements of TF matrices are very different due to the various control modes, the nearly 3 dB difference in the magnitudes of the IMs below $2f_1$, as indicated by the two solid lines, is primarily influenced by the AC/DC dynamics. The dominant peaks observed at $f_1$ are due to the infinite frequency response at $2f_1$ of CCSC while the peaks around $-f_1$ and $-2f_1$ are related to the infinite iteration of modulation of MMCs, which can be confirmed by comparing $\boldsymbol{A}$(n+2, n+2) and $\boldsymbol{BB}_i$(n+2, n+2) in (7).

### B. Mode Positioning

In Fig. 8, using the parameters listed in Table I, the plots of $\text{Re}[D_L(Z^C_{n\_L}(s))]$ and $\text{Im}[D_L(Z^C_{n\_L}(s))]$ for the SE MMC in the CCSC case are shown. The frequency range displayed in the figure spans from $-3f_1$ to $3f_1$ for brevity. The minima observed around $\pm f_1$ and $-2f_1$ of $\text{Im}(\cdot)$ are regarded as zero-pole pairs. Due to the dominance of the pole, the minima of $\text{Im}(\cdot)$ coexist with negative slopes of $\text{Re}(\cdot)$ around the close frequency ranges, it is preliminarily concluded that no RHP zero is identified. This observation aligns with the simulations shown in Fig. 4 before $t$=5 s. A group of Re-Im separation with AC & DC terminal dynamics neglected is replotted in Fig. 8, which indicates the non-negligible effects over $-f_1$ to $2f_1$ and the miss of a mode.

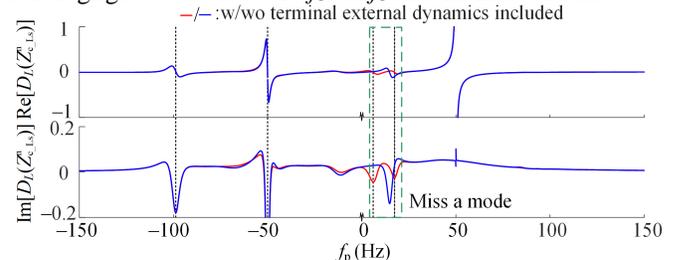

Fig. 8. Re(·) &Im(·) for SE MMC (CCSC case).

Fig. 9 provides further confirmation of the importance of modeling AC and DC terminal dynamics in the inner loop IM. The simulation is conducted for the SE MMC in the FCCC case, using the parameters specified in Table I, except for setting $K_{pov}$ to 1.4. Based on the simulation results shown in Fig. 5 (b) after $t$=35s, the system is expected to be unstable. However, it is observed that only when the terminal dynamics are modeled, a single negative-damping mode around $f_p$=–48Hz can be identified, as indicated by the red solid lines instead of the blue solid lines.

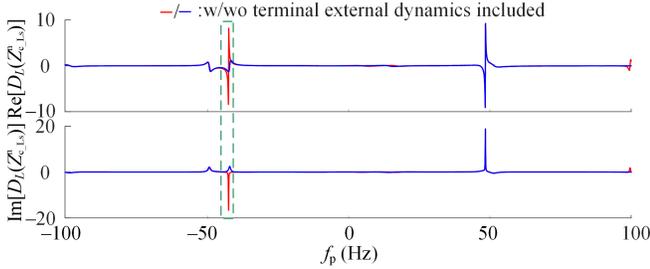

Fig. 9. Re(·) &Im(·) for SE MMC (FCCC case, $K_{pov}$ =1.4).

### C. Validation of Mode Identification

In Figs. 10 and 11, the proposed methods and sensitivity analyses are employed to identify the negative damping modes and provide theoretical explanations for the instabilities observed in Figs. 3 and 4.

Fig. 10 focuses on the CCSC case with RE MMC. The parameters in Table I are kept the same except for $K_{pi2}$. The scan of $K_{pi2}$ is divided into two parts: from 0.05 to 0.3 with a step of 0.05, and from 0.3 to 2.3 with a step of 0.2. For brevity, only 9 groups of plots are shown. Three types of unstable modes are observed with critical values of $K_{pi2}$s between 0.1-0.15 (Fig. 10(a)), 1.3-1.5 (Fig. 10(b)), and 2.1-2.3 (Fig. 10(c)). The identified modes using the unconstrained optimization are as follows:

test 1: $2.04+2\pi\times116.70j$ & $0.25+2\pi\times(-127.83j)$;
test 2: $0.61+2\pi\times(-99.31j)$;
test 3: $3.26+2\pi\times(-556.70j)$ & $0.78+2\pi\times(-99.22j)$.

These identified modes provide accurate explanations for the dominant unstable modes observed in Figs. 3(a)-(c), respectively. It is worth noting that the identified oscillation frequency of the DC variable is given by ($f_p$–$f_1$). A similar process is performed for the SE MMC but no unstable mode is observed. Therefore, it can be concluded that the RE MMC is responsible for the instabilities in the CCSC case, and there are both upper and lower bounds of $K_{pi2}$ that need to be maintained to keep the system stable.

In Fig. 11, the FCCC case with SE MMC is considered. The parameters in Table I are kept the same except for $K_{pov}$, which is varied from 0.1 to 2 with a step of 0.1. For brevity, only 7 groups of plots are shown. Two types of unstable modes are observed with critical values of $K_{pov}$s between 1.3-1.4 and 0.9-1.0 (where the system diverges very slowly when $K_{pov}$=1.0 and is thus neglected). The identified modes using the numerical estimation are as follows:

test 1: $0.13+2\pi\times48.48j$;
test 2: $0.38+2\pi\times(-42.24j)$.

These identified modes provide very accurate explanations for the dominant unstable modes observed in Figs. 4(a)-(b). A similar process is performed for the RE MMC but no unstable mode is observed. Therefore, it can be concluded that the RE MMC is responsible for the instabilities in the FCCC case, and there are both upper and lower bounds of $K_{pov}$ that need to be maintained to keep the system stable.

By combining the results from Figs. 10 and 11, the idea of the *forbidden zone* can be integrated into the graphical assessment to develop the stability margin. The extremums of Im(·) with positive slopes of Re(·) can be limited between 0 and a set positive boundary after the stability enhancements.

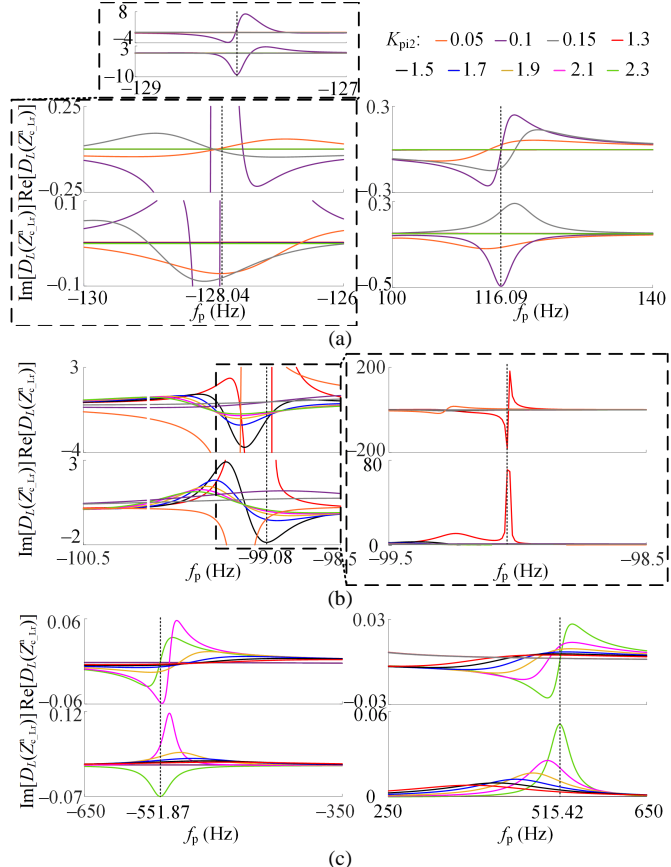

Fig. 10. Sensitivity analysis (CCSC case, RE MMC). (a)-(c) explain tests 1-3.

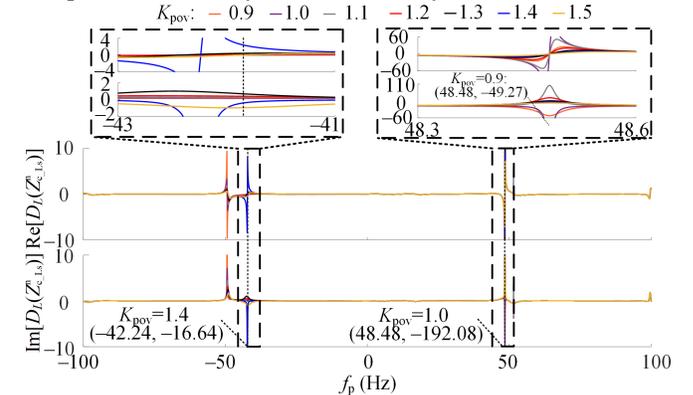

Fig. 11. Sensitivity analysis (FCCC case, SE MMC).

## VI. DISCUSSIONS AND CONCLUSIONS

### A. Remarks on Loop Gain-Based and SS-Based Methods

The case studies in Section V. C can explain the instabilities that the terminal AC/DC stability analysis cannot cover for MMCs, which proves the necessity of improving the impedance-based method in this work. Nevertheless, one should realize that other methods can achieve the same objective and the comments are given below.

In [27], a closed-loop gain is formed using the circulating current reference as the input and the sampled circulating current as the output. This method aims to assess the "self-stability" of an MMC fed by ideal voltage/current sources. While it provides a theoretical basis for CCC design and is believed to reflect the same modes as the inner loop IM with a frequency shift of $2f_1$, it neglects the AC & DC terminal dynamics, making it unable to reveal real modes in interconnected systems, as indicated in Section V.B. In an extreme case [28], an unsatisfactory controller design can be "hidden" by the damping from the grid. Therefore, it may not adapt well to the practical requirements of stability analysis. Additionally, the determination of the number of RHP poles before applying the Nyquist criterion is unclear or requires further operations [27], which can be avoided using the proposed logarithmic derivative-based stability criterion in this work.

SS serves as an important tool for the stability analysis of TSOs in power systems dominated by synchronous generators. Technically, it can be extended to white-box systems with a rational harmonic truncation for MMCs and cover all modes [29]. In terms of the relationship between the proposed method and SS, the discussed inner loop IM can be output by the corresponding SS. However, to the author's knowledge, such progress has not been reported, indicating incompleteness in [4] and [6]. However, even if the proposed loop IM-based method is extended to cover complete physical circuits and identify all modes, it may be challenging to provide reliable in-depth information such as participation factors, as SS does.

Hence, a recommendation for future study is to combine the two methods: using the quantitative information provided by the SS-based method to explain the authentic causes of each unstable mode identified by the loop IM-based method. This is because the approximation of pure time delay in SS introduces many redundant modes, which tends to result in a dimensional explosion for any system-level theoretical analysis involving MMCs.

*B. Transfer Immittance of MMC*

Based on the technical contribution of this work, it is important to extend the concept of transfer immittance, which has been previously applied to consider the MMC [30]-[31] as an AC-DC two-port network similar to TL-VSC [17]. This approach is expected to be well-received in future system-level analyses. For the MMC operating under the active power control mode and being supplied an ideal voltage source, the transfer immittance $Y_{MMC}$ should be a 6×6 matrix:

$$\begin{bmatrix} \Delta i_p^C(s-j3\omega_1) \\ \Delta i_n^D(s-j2\omega_1) \\ \Delta i_z^C(s-j\omega_1) \\ \Delta i_p^D(s) \\ \Delta i_n^C(s+j\omega_1) \\ \Delta i_z^D(s+j2\omega_1) \end{bmatrix} = \begin{bmatrix} Y_{pp}^{CC} & Y_{pn}^{CD} & Y_{pz}^{CC} & Y_{pp}^{CD} & Y_{pn}^{CC} & Y_{pz}^{CD} \\ Y_{np}^{DC} & Y_{nn}^{DD} & Y_{nz}^{DC} & Y_{np}^{DD} & Y_{nn}^{DC} & Y_{nz}^{DD} \\ Y_{zp}^{CC} & Y_{zn}^{CD} & Y_{zz}^{CC} & Y_{zp}^{CD} & Y_{zn}^{CC} & Y_{zz}^{CD} \\ Y_{pp}^{DC} & Y_{pn}^{DD} & Y_{pz}^{DC} & Y_{pp}^{DD} & Y_{pn}^{DC} & Y_{pz}^{DD} \\ Y_{np}^{CC} & Y_{nn}^{CD} & Y_{nz}^{CC} & Y_{np}^{CD} & Y_{nn}^{CC} & Y_{nz}^{CD} \\ Y_{zp}^{DC} & Y_{zn}^{DD} & Y_{zz}^{DC} & Y_{zp}^{DD} & Y_{zn}^{DC} & Y_{zz}^{DD} \end{bmatrix} \begin{bmatrix} \Delta v_p^C(s-j3\omega_1) \\ \Delta v_n^D(s-j2\omega_1) \\ \Delta v_z^C(s-j\omega_1) \\ \Delta v_p^D(s) \\ \Delta v_n^C(s+j\omega_1) \\ \Delta v_z^D(s+j2\omega_1) \end{bmatrix} \quad (17)$$

In (17), the 3×3 submatrix highlighted in red represents the conventional transfer immittance of the MMC, which is not suitable for analyzing the system's internal stability. The relationship between loop IM and transfer immittance is briefly illustrated in Fig. 12. Here, the subscript ′ indicates the influence of other physical circuits apart from the focused MMC, and $Z_{\Sigma g}$ is the equivalent grid side impedance which is a block diagonal matrix. It should be noted that $Z_{\Sigma g}$ is singular because the perturbation $\Delta v_{n/p}^C$ in (17) should be added at the same position of $\Delta v_{ptb}$ in Fig. 1 (which can be any arbitrary position within the circulating circuit).

Using the superposition theorem, each element in (17) can be measured and theoretically deduced in a concise manner, e.g., $Y_{nn}^{CC}$ can be derived by utilizing (7) by blocking the terminal AC & DC dynamics. MMC manufacturers can also adopt a hierarchical stability analysis structure by first using the inner loop IM to identify the partial modes with the support of TSOs as presented in Sections III and IV. Subsequently, the 3×3 submatrix of (17) can be provided to the TSOs to identify the remaining modes involving the subsystem AC-DC interconnection.

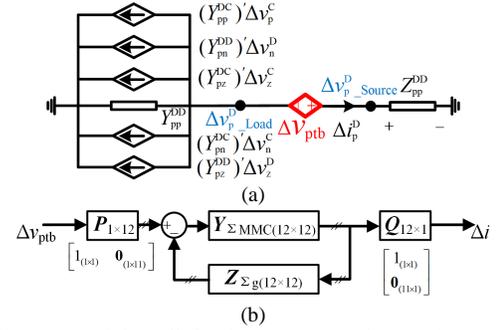

Fig. 12. Illustration of the AC-DM loop IM using the transfer immittance of MMC. (a) Equivalent circuit. (b) Closed-loop representation.

*C. Conclusions*

For a long time, the impedance-based harmonic stability analysis of MMCs lacked an effective tool to identify special modes that can only be observed using the IMs involving the circulating current. This was mainly due to the overextension of experience from TL-VSCs and the incomplete understanding of TF-based analysis from the perspective of classical control theories.

To address this gap, after listing the CCSC and FCCC cases that the terminal AC/DC stability analysis cannot cover, this work provides a comprehensive modeling process of the inner loop IM considering the terminal AC & DC dynamics based on multiharmonic linearization. A logarithmic derivative-based stability criterion is proposed, which allows for the direct identification of modes by factoring out the numerator polynomial of the impedance matrix. This criterion only requires frequency responses and excludes the influences of poles. The presented case studies highlight the importance of control time delay in CCC responses for stability, and MMC manufacturers MUST consider this aspect during controller design. Additionally, by extending the concept of loop IM to all physical circuits in the system, TSOs can theoretically locate oscillations without relying on eigenvalue calculations or superficial simulations.

It is important to note that performing a SISO analysis with the loop IMs deviates from the fundamental idea of using MIMO open-loop gain to assess the stability of closed-loop systems with terminal IMs [9]. However, the SISO analysis is particularly suitable for the MMC-related issues discussed in this work for two reasons: a) achieving a rational source-load partition is challenging, and b) the AC-CM circuit with active control introduced by CCC is an inherent component. The need for information exchange between TSOs and MMC manufacturers, as well as the idea of ensuring internal stability using multiple IMs, should be given attention. In the future, the proposed methods will be extended to more general systems, such as unbalanced or hybrid AC/DC grids, to further enhance their applicability.

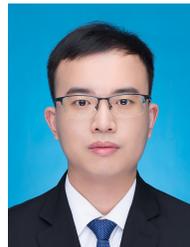

**Chongbin Zhao** (Student Member, IEEE) received the B.S. degree in electrical engineering from Tsinghua University, Beijing, China, in 2019, where he is currently working towards the Ph.D. degree. He is also a visiting scholar at Rensselaer Polytechnic Institute, Troy, NY, United States, started from January to July 2023. His research interests include power quality analysis and control, and emerging converter-driven power system stability analysis and control.

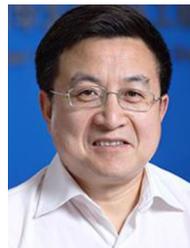

**Qirong Jiang** received the B.S. and Ph.D. degrees in electrical engineering from Tsinghua University, Beijing, China, in 1992 and 1997, respectively. In 1997, he was a Lecturer with the Department of Electrical Engineering, Tsinghua University, where he later became an Associate Professor in 1999. Since 2006, he has been a Professor. His research interests include power system analysis and control, modeling and control of flexible ac transmission systems, power-quality analysis and mitigation, power-electronic equipment, and renewable energy power conversion.